\documentclass[twocolumn,noshowpacs,preprintnumbers,amsmath,amssymb,prl,superscriptaddress]{revtex4}
\usepackage{graphicx}
\usepackage{dcolumn}
\usepackage{bm}
\usepackage{color}

\begin{document}


\title{Puddle-induced resistance oscillations in the breakdown of the graphene quantum Hall effect}

\author{M. Yang}
\affiliation{Laboratoire National des Champs Magn\'etiques Intenses, INSA UPS, CNRS UPR 3228,
Universit\'e de Toulouse, 143 avenue de Rangueil, 31400 Toulouse, France}

\author{O. Couturaud}
\affiliation{Laboratoire Charles Coulomb (L2C), UMR 5221 CNRS-Universit\'e de Montpellier, Montpellier, F-France.}

\author{W. Desrat}

\author{C. Consejo}
\affiliation{Laboratoire Charles Coulomb (L2C), UMR 5221 CNRS-Universit\'e de Montpellier, Montpellier, F-France.}

\author{D. Kazazis}
\affiliation{Centre de Nanosciences et de Nanotechnologies, CNRS, Univ. Paris-Sud, Universit\'e Paris-Saclay, C2N – Marcoussis, 91460 Marcoussis, France}
\affiliation{Laboratory for Micro and Nanotechnology, Paul Scherrer Institute, 5232 Villigen-PSI, Switzerland.}

\author{R. Yakimova}
\author{M. Syv\"aj\"arvi}
\affiliation{Department of Physics, Chemistry and Biology, Link\"oping University, SE-58183 Link\"oping, Sweden}


\author{M. Goiran}
\affiliation{Laboratoire National des Champs Magn\'etiques Intenses, INSA UPS, CNRS UPR 3228,
Universit\'e de Toulouse, 143 avenue de Rangueil, 31400 Toulouse, France}

\author{J. B\'eard}
\affiliation{Laboratoire National des Champs Magn\'etiques Intenses, INSA UPS, CNRS UPR 3228,
Universit\'e de Toulouse, 143 avenue de Rangueil, 31400 Toulouse, France}

\author{P. Frings}
\affiliation{Laboratoire National des Champs Magn\'etiques Intenses, INSA UPS, CNRS UPR 3228,
Universit\'e de Toulouse, 143 avenue de Rangueil, 31400 Toulouse, France}

\author{M. Pierre}
\affiliation{Laboratoire National des Champs Magn\'etiques Intenses, INSA UPS, CNRS UPR 3228,
Universit\'e de Toulouse, 143 avenue de Rangueil, 31400 Toulouse, France}

\author{A. Cresti}
\affiliation{Universit\'e Grenoble Alpes, IMEP-LAHC, F-38000 Grenoble, France}
\affiliation{CNRS, IMEP-LAHC, F-38000 Grenoble, France}

\author{W. Escoffier}
\affiliation{Laboratoire National des Champs Magn\'etiques Intenses, INSA UPS, CNRS UPR 3228,
Universit\'e de Toulouse, 143 avenue de Rangueil, 31400 Toulouse, France}

\author{B. Jouault}
\affiliation{Laboratoire Charles Coulomb (L2C), UMR 5221 CNRS-Universit\'e de Montpellier, Montpellier, F-France.}

\begin{abstract}
We report on the stability of the quantum Hall plateau in wide Hall bars made from a chemically gated graphene film grown on SiC. The $\nu=2$ quantized plateau appears from fields $B \simeq 5$ T and persists up to $B \simeq 80$ T. At high current density, in the breakdown regime, the longitudinal resistance  oscillates with a $1/B$ periodicity and an anomalous phase, which we relate  to the presence of additional electron reservoirs. 
The high field experimental data suggest 
that these reservoirs induce a continuous increase of the carrier density up to the highest available magnetic field, thus enlarging the quantum plateaus. These in-plane inhomogeneities, in the form of high carrier density graphene pockets, modulate the quantum Hall effect breakdown and decrease the breakdown current.
\end{abstract}

\maketitle
Graphene~\cite{graphene,Zhang2005Nature} 
shows a unique half-integer quantum Hall effect (QHE) with conductivity plateaus
$\sigma_{xy}= 4(m+1/2)e^2/h$ where the factor 4 stands for the spin and valley degeneracies
and $m= 0, \pm 1, \pm 2...$~\cite{GoerbigRMP2011}.
The peculiar dispersion
$E_m \simeq \pm 420 \sqrt{|m| B [\mathrm{T}]}$ K
of the Landau levels (LLs) in graphene induces large energy gaps at low LL indices
and allows to explore exotic transport phenomena
even at relatively high temperature~\cite{Novoselov1379}.
The substrate graphene is deposited on plays a central role in determining the features of the observed QHE. 
In low-mobility graphene on SiO$_2$, the presence of electron-hole puddles at the charge neutrality point (CNP) prevents any divergence of the longitudinal resistance at filling factor $\nu < 2$, whereas the Hall resistivity fluctuates around zero due to charge compensation \cite{Poumirol2010PRB}. On the other hand, in high mobility graphene deposited on boron nitride flakes, a complete degeneracy lifting of the Landau levels can be observed and the corresponding spin-valley textures have been identified \cite{Bolotin2009nature,Dean2011NatPhys}.
In this Letter, we consider graphene deposited on top of a SiC substrate (G/SiC).
In this system, the quantum Hall plateau at $h/2e^2$ is exceptionally robust with respect to magnetic field~\cite{Alexander2013PRL}. 
Moreover, G/SiC shows metrological quantum Hall quantization, with relative accuracies of the quantized resistance better than $10^{-9}$\cite{Tzalenchuk2010Nature},
even at lower magnetic fields and higher temperatures than GaAs-based quantum Hall resistance standards~\cite{Ribeiro}.
In this context,
it is important to unveil the role of charge reservoirs in the stabilization
of the first quantum Hall plateau~\cite{Zawadzki,Janssen2011PRB} and
to identify the mechanisms governing the breakdown of the QHE~\cite{Nachtwei1999,Alexander2013PRL,Lafont2015Nature}.

\begin{figure}
\includegraphics[width=0.95 \linewidth]{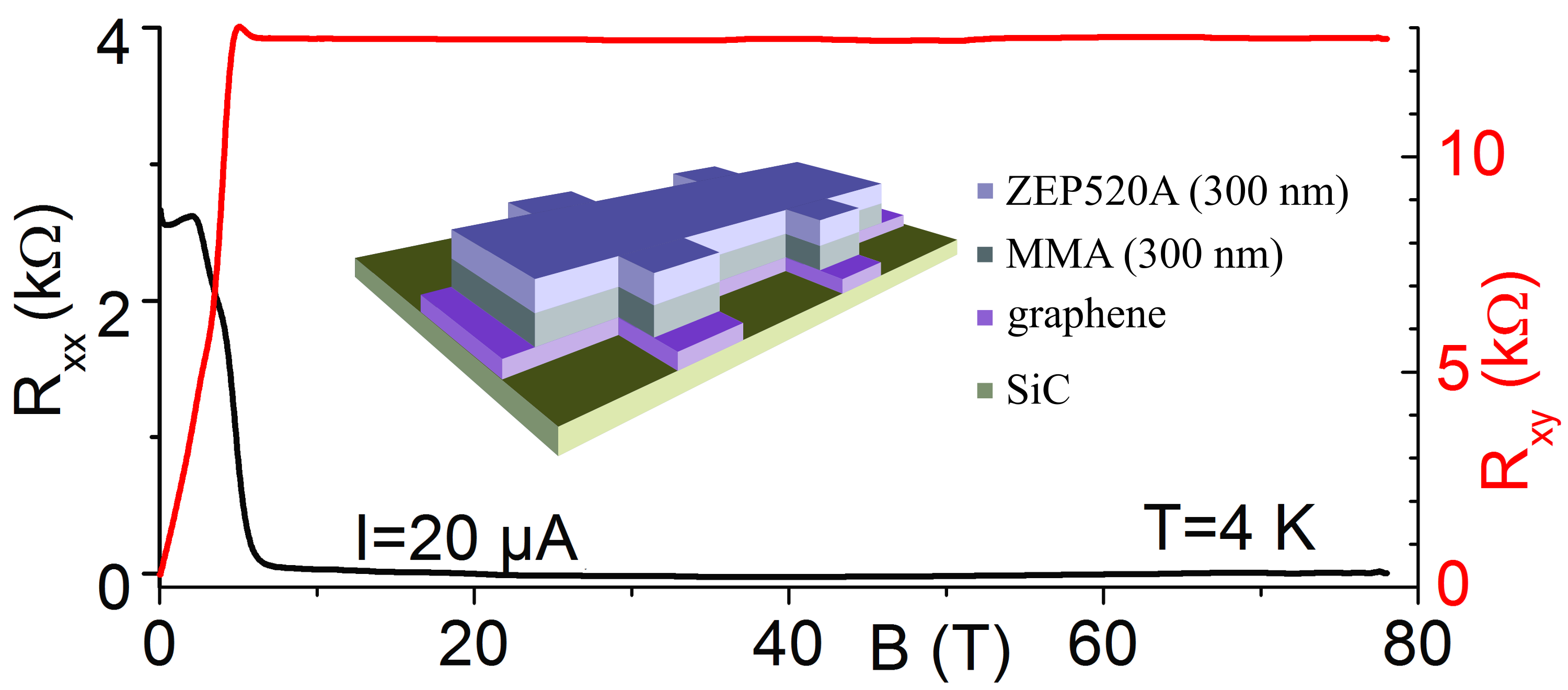}
\caption{(a) Longitudinal and Hall magnetoresistances of sample S1 measured at $T$= 4 K up to $B$= 80 T. The inset is a sketch of the Hall bar covered by the resists.}
\label{fig:QHE80T}
\end{figure}

In the following, we report on the results obtained from graphene samples
grown by “Graphensic AB” company on the Si-face of a 4H-SiC substrate.
The details concerning the growth conditions can be found in Ref.~\onlinecite{Tzalenchuk2011SSC}.
The samples are tailored into Hall bar geometry with a width  $W= 100$~$\mu$m and a total length of $420$~$\mu$m.
The as-grown carrier density of G/SiC is of the order of $1\times10^{13}$ cm$^{-2}$ 
and is reduced to about $\simeq 5\times10^{11}$ cm$^{-2}$ using a polymer gate (photo-chemical doping) \cite{Tzalenchuk2011SSC}. The mobility at $T =  4$ K is about 7,000 cm$^2$V$^{-1}$s$^{-1}$. 
The Hall bars are electrically connected using wedge bonding and placed in a low temperature and high (pulsed) magnetic field setup for magneto-transport measurements. 

Figure \ref{fig:QHE80T} shows the longitudinal and Hall resistances $R_{xx}$ and $R_{xy}$ of
one sample, named S1, measured as a function of the magnetic field at a temperature $T=4.2$~K. 
In this Letter, the longitudinal resistance is normalized to a square. 
For $B \ge 7$ T, an extremely wide quantum Hall plateau at $R_{xy}=h/2e^2$ is observed up to the highest available magnetic field (78 T).  
As expected, the quantum Hall plateau coincides with a complete vanishing of $R_{xx}$. 
\begin{figure}[h]
\includegraphics[width=1.0 \columnwidth]{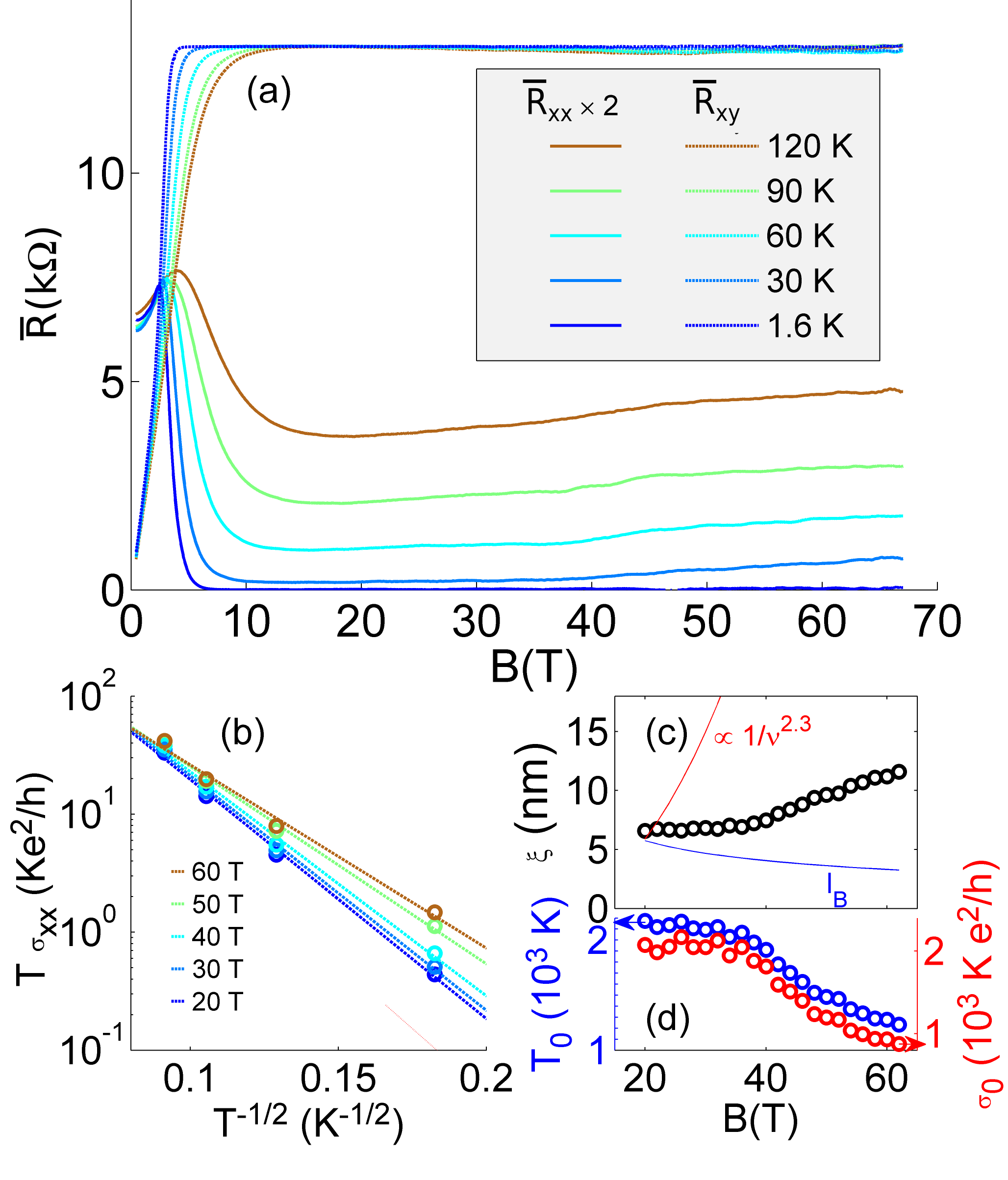}
\caption{(a) Longitudinal (solid lines) and transverse (dashed lines) symmetrized resistances for sample S1, at $T=1.6$ K up to $T=120$ K. 
(b) solid lines, open symbols: $T \sigma_{xx}$ {\it vs} $T^{-1/2}$ from $B$= 20 T up to $B$= 60 T with a step of 10 T. Dashed lines: linear fits with the soft Coulomb gap VRH model. The data of Ref.~\onlinecite{Lafont2015Nature} at $B=19$ T are reported for comparison (red dashed line). (c) Open symbols: localization length $\xi$ extracted from the VRH fit.  The blue and red lines are guides for the eye, following the magnetic length $l_B$ and the prediction $\xi \sim 1/\nu^{2.3}$ respectively. (d) Parameters $T_0$ and $\sigma_0$  extracted from the fit.}
\label{fig:S1VRH}
\end{figure}
A small mixing between $R_{xx}$ and $R_{xy}$, due to the anisotropy of the conductivity induced by the SiC steps~\cite{Schumann2012PRB,Pallecchi2014SR}, is noticed as a small additional bump in $R_{xy}$ at $B \simeq$ 7 T.
This mixing can been 
removed  
by (anti)symmetrizing the measured resistances recorded in the two opposite directions of the magnetic field:
$\bar{R}_{xx(xy)}= [R_{xx(xy)}(B) \pm R_{xx(xy)}(-B)]/2$. 
Figure~\ref{fig:S1VRH}(a) shows the symmetrized magnetoresistances of sample S1 at different temperatures.
The longitudinal resistance $\bar{R}_{xx}$ has a minimum around $B_\mathrm{min} \simeq 18$ T when $T$ increases above 30~K.
We assume that, as in two-dimensional semiconductor systems, 
this minimum occurs when the Fermi energy is at mid-gap between two adjacent Landau levels.
This implies that the filling factor $\nu$ equals $2$ at $B=B_\mathrm{min}$ and
the  carrier density increases
from  $3.4 \times 10^{11}$  cm$^{-2}$ at $B=0$ T
up to $8.5\times 10^{11}$ cm$^{-2}$ at $B=B_\mathrm{min}$.

This phenomenon is in qualitative agreement with the model proposed in Ref.~\cite{Kopylov2010APL}, based on a charge transfer taking place between graphene, the chemical gate and the interface states between SiC and graphene. 
In particular, it was shown~\cite{Janssen2011PRB,Alexander2013PRL} that the carrier
concentration in similar systems can increase by 200 \% from $B=0$ T to $B=16$ T. However, at higher magnetic fields, when only the $m=0$ LL located at the CNP is populated, the model predicts 
the electron concentration saturation, which is hardly
compatible with our experimental findings. Indeed, by assuming a constant carrier density $n=8.5 \times 10^{11}$ cm$^{-2}$ above $B_\mathrm{min}$, we would have $\nu=0.45$ at $B=78$ T. Therefore, the large plateau observed in Fig. 1 questions the charge transfer model and deserves further attention.

At low temperatures, the electron diffusion in the quantum Hall regime is often well described 
by the variable range hopping (VRH) theory in the soft Coulomb gap regime\cite{VRH},
where the conductivity reads:
\begin{equation}
\sigma_{xx} = \frac{\sigma_0}{T} \exp(-\sqrt{\frac{T_0}{T}}).
\label{eq:vrh}
\end{equation}
Here, $\sigma_0$ is a parameter,
$T_0$ is the hopping temperature, which is related to the
localization length $\xi$ by the relation
$
k_B T_0 = C e^2 / 4 \pi \epsilon_r \epsilon_0 \xi,
$
$\epsilon_0$ is the vacuum dielectric constant,
$\epsilon_r \simeq 7$ is the relative dielectric constant
averaged between the dielectric constants of the resist and SiC,
and $C \simeq $ 6.2 is a constant\cite{Lien1984, Furlan1998PRB}.
VRH has already been observed in various graphene samples below $T=100$ K \cite{Giesbers2009PRB,Bennaceur2012PRB,Janssen2013RPP,Lafont2015Nature}.
The analysis of sample S1, shown in Fig.~\ref{fig:S1VRH}(b), 
uses only the data in the range $T\ge 30$ K, where measurement errors are negligible, and $T \le 120$ K, where conduction by activation to the extended states can be neglected~\cite{suppinfo}.
The data are satisfactorily fitted by Eq.~\ref{eq:vrh} over two decades.
The localization length $\xi$, 
the $\sigma_0$ coefficient
 and the hopping temperature $T_0$ are extracted from the VRH fits,
see Fig.~\ref{fig:S1VRH}(c,d). At low magnetic fields $B \simeq 20$ T, $\xi$ is comparable to the magnetic length $l_B = \sqrt{\hbar/eB}$. At higher $B$, $\xi$  increases only slightly.
In the QHE regime, the localization length  is expected to vary according to
$\xi \propto  \Delta E^{-\gamma}$, 
where $\Delta E$ is the mobility gap between the nearest delocalized states and the Fermi energy and
$\gamma \simeq 2.3$ is the quantum percolation exponent~\cite{Huckestein}.
Assuming 
a constant charge density $n$ and
a density of localized states independent of energy, 
the localization length
$\xi^\mathrm{th}_B$ would vary as $1/\nu^{2.3} \sim B^{2.3}$, yielding
$\xi^\mathrm{th}_{B=60\mathrm{T}}/\xi^\mathrm{th}_{B=20\mathrm{T}} \simeq 12$. 
By contrast, the experimental variation
of $\xi$ in the 20--60 T range is about
$\xi_{B=60\mathrm{T}}/\xi_{B=20\mathrm{T}} \simeq 2$, see Fig.~\ref{fig:S1VRH}(c). 
This gives $n_{B=60\mathrm{T}} \simeq 2.2 \times n_{B=20\mathrm{T}}$, thus indicating
a carrier density increase even for $B > B_\mathrm{min}$.
%

\begin{figure}
\includegraphics[width=1.0 \columnwidth]{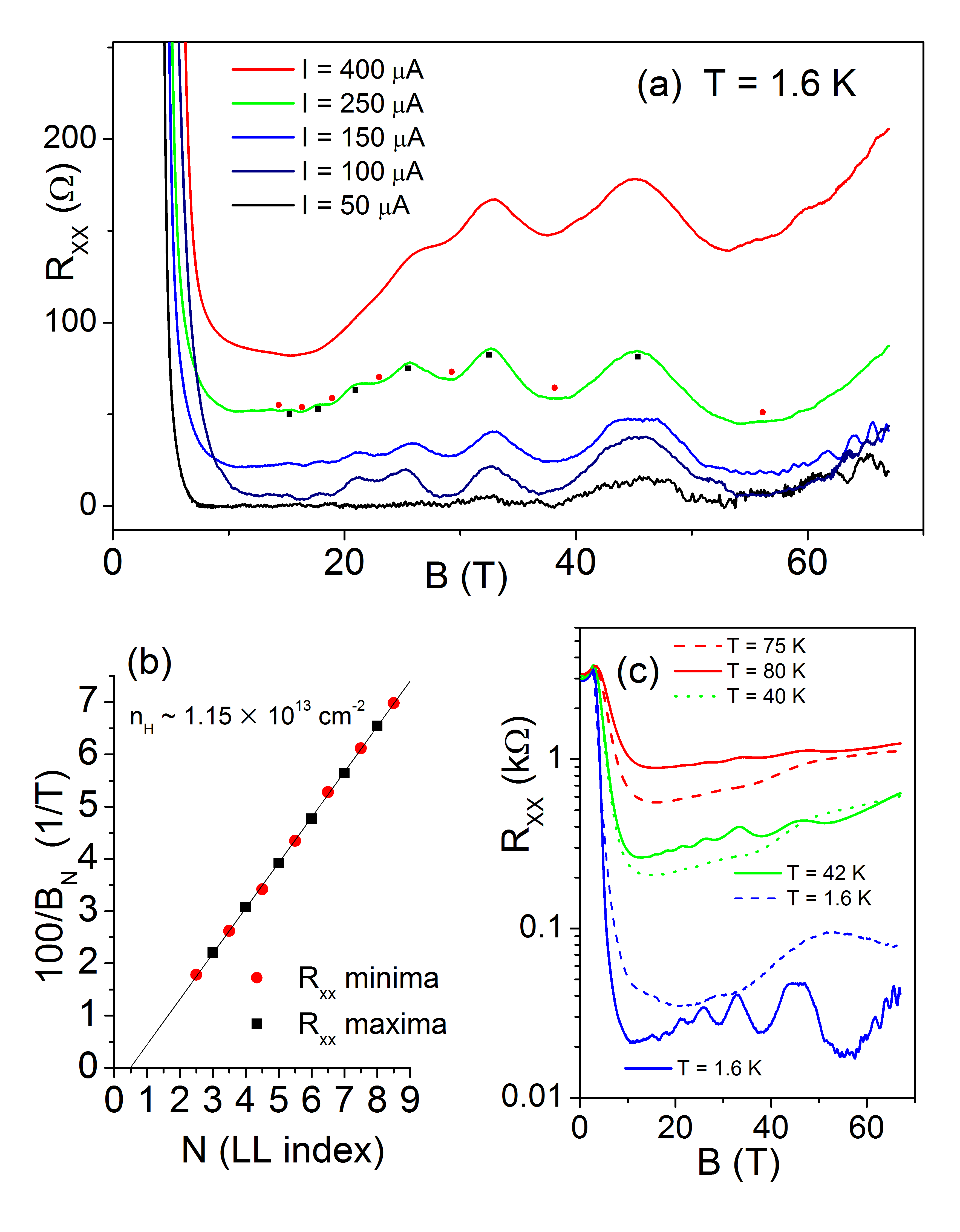}
\caption{(a) Longitudinal resistance $R_{xx}$ for currents higher than the breakdown current $I_c \simeq$ 10 $\mu$A. $1/B$-periodic oscillations are visible.  (b) Landau plot for the minima and maxima of these oscillations. The period corresponds to a carrier concentration $n_\mathrm{HD} \approx 1.15 \times 10^{13}$ cm$^{-2}$. The $R_{xx}$maxima plotted as a function of their index $N$ go to $N=1/2$ when $B_N \rightarrow \infty$, which is usually a signature of a zero Berry Phase. (c) The observed oscillation damping by the temperature does not agree with the usual
Shubnikov--de Haas theory. The oscillation period may change from one cool-down (solid line) to another
(dashed lines).}
\label{fig:S1current}
\end{figure}

Figure~\ref{fig:S1current} shows $R_{xx}$ of sample S1 for current $I$ in the range $50$--$400$ $\mu$A.
At $I= 100$ $\mu$A,  $R_{xx} > 3~\Omega$ over the whole magnetic field range, 
yielding a longitudinal voltage ($V_{xx} > 300$~$\mu$V) too large to be
compatible with metrological measurements.
This defines a critical breakdown current $I_c \simeq 1$ A/m,
one order of magnitude lower than the highest reported values for graphene on SiC~\cite{Alexander2013PRL}.
Moreover, $R_{xx}(B)$ does not have any more a unique minimum associated to $\nu=2$ but
shows oscillations periodic in $1/B$. 
The periodicity is evidenced in the Landau plot reported in Fig.~\ref{fig:S1current}(b).
Assuming a degeneracy of spin and valley, the period $\Delta(1/B)$ of these oscillations corresponds
to a constant high carrier density $n_\mathrm{HD} = 4 (e/h) / \Delta(1/B)$ $\simeq$ 1.15 $\times 10^{13}$ cm$^{-2}$.
 
Similar $1/B$-periodic oscillations  have been observed in other samples.
The periodicity corresponds to concentrations in the range $4-12 \times 10^{12}$ cm$^{-2}$. 
The period and amplitude of the oscillations change each time a thermal cycle takes place
but the period does not change when the current is increased.
In all of the samples, 
resist cracks
appeared
after several weeks 
of measurements at low temperatures and several temperature cycles.
We postulate that 
even before the appearance of cracks, the chemical gating is not homogeneous
and microscopic detachments of the resist
lead to disconnected graphene puddles with native High carrier Density (HD), 
whereas the rest of the Hall bar
remains covered by the resist and has
a Low electron Density (LD).
Additionally, 
graphene detachment from the substrate induced by the SiC steps~\cite{LowPRL2012},
bilayer patches~\cite{panchal20162DM} or residual contamination~\cite{erikssonAPL2012}
may also induce inhomogeneous carrier density.

For most samples, 
the $I/V_{xx}$ characteristic is ohmic,
the $1/B$ oscillations appear independently of the injected current
and the transport is dissipative.
This is not the case for sample S1,  which shows a more remarkable behavior.
There, the oscillations appear only above a critical current $I_c \simeq 10$ $\mu$A 
in the breakdown regime of the quantum Hall effect.
At lower current, S1 keeps good metrological properties,
as evidenced by additional precision measurements
performed in constant magnetic field.
At $B=10$~T, $T=1.8$ K and a DC current $I= 10$ $\mu$A
the deviation of the Hall resistance from the expected value $h/2e^2$ is
$(\Delta R_H - h/2e^2)/(h/2e^2)$ =  $(-0.4 \pm 10) \times 10^{-6}$,
while  the longitudinal resistance is $(10 \pm 17)$ m$\Omega$.
This precision and this uncertainty correspond to the limit of the setup.
This analysis rules out parallel conduction.
Another key feature of such $1/B$-periodic oscillations for all samples 
is that their Berry phase is zero, as evidenced by the Landau plot of Fig.~\ref{fig:S1current}(b) for sample S1. 

We can take advantage of these unusual features to 
unravel the role of the inhomogeneity in the carrier density in the stabilization of the quantum Hall plateau.
We call $S^\mathrm{HD}$ and $S^\mathrm{LD}$
the total areas corresponding to the HD and LD regions.
We label $n^\mathrm{i}_0$ and $\mu^\mathrm{i}_0$ 
the charge density and the chemical potential in each region $i \in \{\mathrm{HD},\mathrm{LD}\}$ in the absence
of magnetic field.
Here $\mu^i_0$ is the chemical potential with the energy reference at the CNP of region $i$.
These parameters are constant:
$n^\mathrm{HD}_0\simeq 11.5 \times 10^{12}$ cm$^{-2}$ 
from the $1/B$-period of the oscillations and
$n^\mathrm{LD}_0=10^{11}$ cm$^{-2}$ from the Hall effect at low field $B \simeq 0.1$ T.
We label $\Delta\mu^\mathrm{HL}$= $\mu^\mathrm{HD}_0 - \mu^\mathrm{LD}_0$
= 330~meV the energy difference between the CNP points in the LD and HD regions.

\begin{figure}[h]
\includegraphics[width=1.0 \columnwidth]{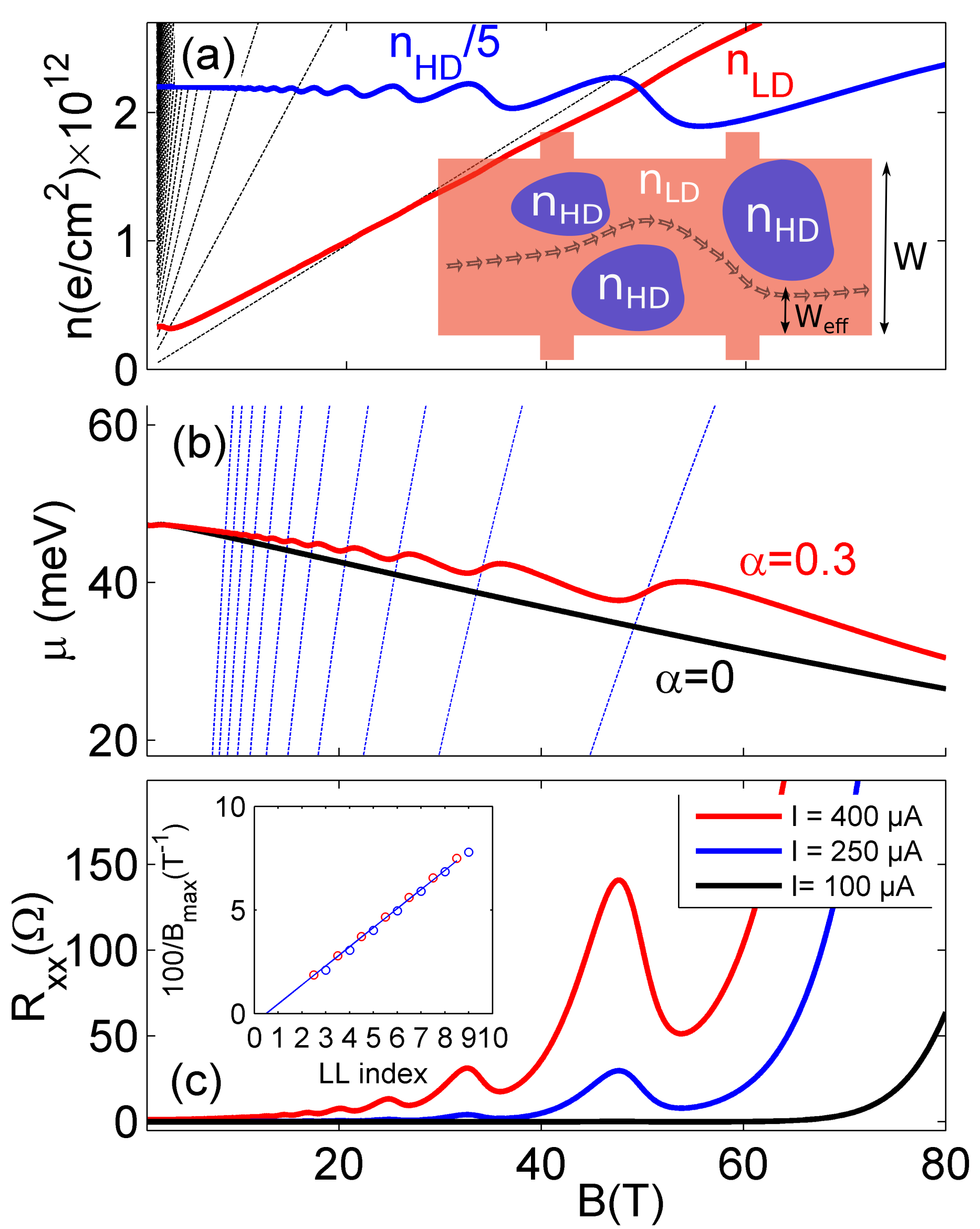}
\caption{(a) Charge density in LD and HD regions as a function of the magnetic field. The proportion of highly doped regions $\alpha$ is fixed to 30\% of the total surface (thick blue and red lines: $n_\mathrm{HD}$ and $n_\mathrm{LD}$ respectively), $A=0.4$ eV and $\Gamma= 15$ meV. The thin dashed lines represent the LL carrier density in LD region. The inset is a sketch of the inhomogneous Hall bar embedding several highly doped regions, a deflected current line is sketched by open arrows.
(b) Chemical potential $\mu$ as a function of $B$ for $\alpha$= 30\% (red line) and $\alpha=0$ (black line). The thin blue dashed lines represent the positions of the Landau Levels in the HD puddles. 
(c) Longitudinal resistance $R_{xx}$ as a function of $B$, at various currents, calculated within the model described in the text. The curves from bottom to top correspond to currents of 100, 250 and 400 $\mu$A respectively. The inset is a Landau plot (blue circles: maxima indexed with integers, red circles: minima indexed with half-integers).
}
\label{fig:ct}
\end{figure}

According to Ref.~\onlinecite{Kopylov2010APL},
for each homogeneous region $i$ the electron density is given by
$
n^{i}=-n_{g}^{i}+\beta (A-\mu^i)
\label{eq3}
$
where 
%
$n_g^{i}$ is the electron density pinned by the electrochemical gate,
$A$ is the difference between the work function of undoped graphene and donor states
and $\beta$ is an effective density of states.
In the highly doped region,
no polymer is present and the charges come from the substrate.
Therefore $n_{g}^\mathrm{HD}=0$
and $\beta= n^\mathrm{HD}_0/ (A- \mu^\mathrm{HD}_0)$.
In the low doped region, the polymer acts as a gate 
and then $n_g^\mathrm{LD}= -n^\mathrm{LD}_0 + 
n^\mathrm{HD}_0 (A-\mu^\mathrm{LD}_0)/(A-\mu^\mathrm{HD}_0)$.
Once these parameters have been determined, we let the magnetic field change.
The carrier density in each region evolves
according to the dispersion of the graphene LLs 
characterized by energies $E_m$ and a Lorentzian broadening $\Gamma$. 
Since the regions are coupled, electrons redistribute to keep 
the same chemical potential $\mu$ everywhere: 
$\mu= \mu^\mathrm{LD} = \mu^\mathrm{HD}-\Delta\mu^\mathrm{HL}$.
For each value of the magnetic field, $\mu$ is determined by solving numerically
$\bar{n}= (1-\alpha)\left[ -n_g^\mathrm{LD}+ \beta (A-\mu)\right] + \alpha \beta \left[A- \mu -\Delta\mu^\mathrm{HL} \right]$
where $\bar{n}$ is the average electron concentration:
$\bar{n}= (1-\alpha) n^\mathrm{LD}+  \alpha n^\mathrm{HD}$
and
$\alpha= S^\mathrm{HD}/(S^\mathrm{LD}+S^\mathrm{HD})$.
The only remaining fitting parameters are $A$, $\Gamma$ and $\alpha$.
We determine the magnetic field-dependent electron density in the two regions as shown in Fig.~\ref{fig:ct}(a). 
The charge density in LD region  increases without saturation up to high magnetic field whereas it oscillates in HD region.
These oscillations are $1/B$-periodic and in turn induce a modulation of the filling factor with an {\it opposite phase} in the LD region. In the transport properties, this phenomenon translates into $1/B$-periodic oscillations of $R_{xx}(B)$ with zero apparent Berry phase each time the Fermi energy approaches the LL extended states.

Each HD puddle, with a conductivity much higher than the conductivity of the LD region,
behaves as an equipotential region.
The direction of the electric field at the HD/LD interface is normal to the interface.
In sufficiently high magnetic fields,
the current forms a Hall angle close to 90$^o$ with the electric field
and cannot enter into
the HD puddles which act as electrically insulating regions.
This implies 
a dramatic reduction of the effective Hall bar width $W_\mathrm{eff}$
over which the current flows.
In the VRH theory~\cite{PolyakovPRL1993}, the current acts as an effective temperature
$T_\mathrm{eff}$ given by the relation:
\begin{equation}
T_\mathrm{eff}= e R_H I \xi / (2 k_B W_\mathrm{eff}).
\label{eq:teff}
\end{equation}
Experimentally,
$T_\mathrm{eff}$ is determined by matching the conductivities $\sigma_{xx}(I) = \sigma_{xx}(T_\mathrm{eff})$.
At $B < 20$ T, where the $1/B$-periodic oscillations are small, 
$T_\mathrm{eff}$ appears to be roughly linear {\it versus} $I$, in agreement with Eq.~\ref{eq:teff}.
The relation $\sigma_{xx}(I) = \sigma_{xx}(T_\mathrm{eff})$ yields
$T_\mathrm{eff}$ $\simeq$ 0.05--0.1 $\times (I [\mu$A$])$ [K] 
which corresponds to an effective Hall bar width
$W_\mathrm{eff} \simeq 15$ $\mu$m considerably smaller than the nominal Hall bar width (100 $\mu$m).
A similar discrepancy between $W_\mathrm{eff}$ and $W$ was observed in Ref.~\onlinecite{Furlan1998PRB,Lafont2015Nature}, again suggesting a very inhomogeneous current distribution in the samples.

Whereas the bootstrap electron heating theory is most widely accepted to account for the breakdown of the QHE~\cite{Nachtwei1999},
here the underlying VRH mechanism alone is sufficient to explain the observed $R_{xx}(B)$ dependence.
The localization length varies as
$\xi = a \mu^{-\gamma}$, where $a$ is chosen
to have $\xi_{B=60 \mathrm{T}} = 12$ nm, see Fig.~\ref{fig:S1VRH}(c)
and $\mu$ is given by the calculation, 
see Fig.~\ref{fig:ct}(b). 
 Introducing $\xi$ in $T_{\mathrm{eff}}$ and $T_{\mathrm{eff}}$ in Eq.~(\ref{eq:vrh}),
$R_{xx}$ is calculated at various currents and magnetic fields, as shown in Fig.~\ref{fig:ct}(c).
The $1/B$ periodic oscillations are indeed reproduced, with the correct experimental phase and amplitude.
The best agreement with the data is found by choosing $A=0.4$ eV, $\alpha= 30\%$ and $\Gamma= 15$ meV.
The large $\alpha$ value points towards a largely inhomogeneous sample
and is consistent with the small $W_\mathrm{eff}/W$ value.
The determination of a more quantitative relation between $\alpha$ and $W_\mathrm{eff}$
would require a detailed knowledge of the current distribution.   
The value $A = 0.4$ eV is widely accepted in the literature,
as well as the relation 
$\gamma = \beta \epsilon_0 \epsilon_r / (\epsilon_0 \epsilon_r-e^2d\beta)$,
where $\gamma$ is the density of states at the SiC/graphene interface
and  $d$ is the graphene thickness.
However, the condition $\gamma>0$ implies
$\epsilon_r > 130 \times d$[nm]. 
The value $\epsilon_r \simeq 7$
yields $d < 0.05$ nm,
suggesting 
i) a graphene layer closer to the donor interface states than usually assumed;
ii) an enhancement of the dielectric screening by the $\pi$ electrons.
Using the Random Phase Approximation~\cite{GoerbigRMP2011}, one gets $\epsilon_\mathrm{RPA}^\infty \simeq 13$.
 
To conclude, we addressed the robustness of the $\nu=2$ QH plateau for graphene on SiC.
We observed that this plateau can surprisingly persist at least up to 78 T.
Faint and rapid oscillations on top of the longitudinal resistance also appear in the breakdown regime. 
This effect was demonstrated to be intrinsic to graphene and to originate from
non-homogeneous in-plane doping of the device.
A charge transfer~\cite{RN379} between the SiC substrate and 
graphene regions of different carrier concentration 
is entailed by the magnetic field due to a modification of the band structure and the formation of Landau levels. 
These additional charge reservoirs
modulate the breakdown mechanisms,
lower the breakdown current 
and help to stabilize the quantum plateau up to very high fields.
\begin{acknowledgments}
Part of this work was performed at LNCMI under EMFL proposal TSC06-115.
\end{acknowledgments}


\end{document}